\documentstyle[12pt]{article}
\def\beqa{\begin{eqnarray}}
\def\eeqa{\end{eqnarray}}
\def\beq{\begin{equation}}
\def\eeq{\end{equation}}

\begin{document}
\begin{titlepage}
   \title{Neutrino Oscillations in Brans--Dicke Theory of Gravity}
\author{S. Capozziello\thanks{E-mail: capozziello@physics.unisa.it}~~ and G.
Lambiase\thanks{E-mail: lambiase@physics.unisa.it} \\ {\em
Dipartimento  di Scienze Fisiche ``E.R. Caianiello''} \\ {\em
Universit\`a di Salerno, 84081 Baronissi (Sa), Italia.} \\ {\em
Istituto Nazionale di Fisica Nucleare, Sez. di Napoli.} }
\date{\today}
\maketitle
\begin{abstract}
Flavor oscillations of neutrinos are analyzed in the framework of
Brans--Dicke theory of gravity. We find a shift of quantum
mechanical phase of neutrino proportional to $G_N\Delta m^2$ and
depending on the parameter $\omega$. Consequences on atmospheric,
solar and astrophysical neutrinos are discussed.
\end{abstract}

\thispagestyle{empty}

\vspace{20.mm}


\vfill

\end{titlepage}

\section{Introduction}
\setcounter{equation}{0}

Among all alternative theories of gravity, the Brans--Dicke (BD)
theory \cite{brans} provides the most natural generalization of
General Relativity. It can be thought of as a minimal extension of
Einstein theory in which Mach's principle and Dirac's large number
hypothesis (see, for example, \cite{weinberg}) are properly
accommodated by means of a nonminimal coupling between the
geometry and a scalar field $\phi$, the BD scalar. The scalar
field rules dynamics together with geometry and, furthermore,
induces a variation of the gravitational {\it coupling} with time
and space through the relation $G_{eff}=1/\phi$. The gravitational
constant $G_N$ is recovered in the limit $\phi\to constant$. Some
recent experiments \cite{pioneer} seem to confirm a variation of
the Newton constant on astrophysical and cosmological sizes and
time scale.

The effective action describing the interaction of the scalar
field $\phi$ nonminimally coupled with the geometry and the
ordinary matter is given by \cite{brans}
\begin{equation}\label{1.1}
{\cal A}=\int d^4x \sqrt{-g}\left[\phi R-\omega
\frac{\partial_{\mu}\phi\partial^{\mu}\phi}{\phi}+\frac{16\pi}{c^4}
{\cal L}_{m}\right]\,,
\end{equation}
where $R$ is the scalar curvature, ${\cal L}_{m}$ is the matter
contribution in the total Lagrangian density. The constant
$\omega$ is determined by observations and its value can be
constrained by classical tests of General Relativity. The
consequences of BD action (\ref{1.1}) have been analyzed for the
light deflection, the relativistic perihelion rotation of Mercury,
and the time delay experiment, resulting in reasonable agreement
with all available observations thus far provided $\omega \geq
500$ \cite{will}. On the other hand, bounds on the anisotropy of
the microwave background radiation give the upper limit
$\omega\leq 30$ \cite{la}. Einstein's theory is recovered for
$\omega\to\infty$. In this limit, the BD theory becomes
indistinguishable from General Relativity in all its predictions.

Understanding if the BD theory of gravity may be considered as the
right generalization of Einstein gravity and, as a consequence,
how it affects physical phenomena is an important matter. In this
paper we will face this issue by considering neutrino
oscillations, calculating, in particular, the contribution to the
quantum mechanical phase mixing induced by the non--standard
coupling between the geometry and the scalar field. As we will
see, such a correction does depend on the value of the parameter
$\omega$.

It is well known that the problem of neutrino oscillations is
still open, and the research of new effects in which they could
manifest is one of the main task of the today physics. For this
reason, the quantum mechanical phase of neutrinos propagating in
gravitational field (usually the Schwarzschild or Kerr field) has
been recently discussed by several authors (see
\cite{burgard}--\cite{capozziello} and references therein), also
in view of astrophysical consequences.

More controversial is the debate concerning the red-shift of
flavor oscillation clocks, given by a term proportional to
\begin{equation}\label{1.2}
\frac{G_N\Delta m^2 M}{\hbar E}\, \log\frac{r_B}{r_A}\,,
\end{equation}
first derived by Ahluwalia and Burgard \cite{burgard} in the
framework of the weak gravitational field of a star, with mass
$M$. Here $\Delta m^2$ is the mass--squared difference, $\Delta
m^2=\vert m_2^2-m_1^2\vert$, $E$ the neutrino energy, $r_A$ and
$r_B$ the points where neutrinos are created and detected,
respectively. They also suggest that the oscillation phase
(\ref{1.2}) might have a significant effect for supernova
explosions due to the extremely large fluxes of neutrinos produced
with different energies, corresponding to the flavor states.

This result has been confirmed in the paper by Grossman and Lipkin
\cite{grossman}, and it has been also derived by Konno and Kasai
\cite{konno} under the assumption that the radial momentum of
neutrinos is constant along the trajectory of the neutrino,
strengthening the correctness of the Ahluwalia--Burgard arguments.
Nevertheless, assuming that the neutrino energy is constant along
the trajectory, Konno and Kasai show that the term (\ref{1.2}) is
cancelled out, recovering in such a way the result of Refs.
\cite{giunti,mottola}.

Without pretending to solve or face here this controversy, which
goes beyond our aim, this paper is a straightforward extension of
the calculations of Ref. \cite{giunti} in the framework of BD
theory. In \cite{giunti}, the neutrino oscillation formula in a
gravitational field is based on the {\it covariant form} of the
quantum phase that arises due to the assumed mixing of massive
neutrino. The result (i.e. the cancellation of $G_N\Delta m^2$
term) is the same of Ref. \cite{mottola}, but it is derived
without invoking the assumption that underlying mass eigenstates
are emitted at different time. We find that the scalar field in
(\ref{1.1}) nonminimally coupled to the scalar curvature induces a
red-shift of flavor oscillation clocks in the quantum dynamical
phase, which is proportional to
\begin{equation}\label{1.3}
\frac{G_N\Delta m^2}{E}\frac{1}{2+\omega}\,\log\frac{r_B}{r_A}\,.
\end{equation}
It vanishes in the limit $\omega\to 0$\footnote{The extension of
the paper \cite{burgard} (or \cite{grossman}) to the BD theory
does not give appreciable correction to the quantum dynamical
phase. In fact, the corrective factor is of the form
 \begin{equation}\label{1.4}
 \frac{3+2\omega}{2(2+\omega)},
 \end{equation}
 which is $\sim 1$ for $\omega \sim 500$ and $\omega \leq 30$.}.
 Eq. (\ref{1.3}) can be seen, in some sense, as a further test, in
addition to the standard ones above discussed, for establishing
the validity (or not) of the BD theory.

The layout of this paper is the following. In Sect. 2 we shortly
recall the Schwarzschild--like solution coming from BD field
equations, which describe the static and stationary gravitational
field generated by a mass $M$, and the corresponding expressions
in the weak field approximation (for details, see the paper
\cite{brans}). Sect. 3 is devoted to the calculation of the
quantum mechanical phase for propagating neutrinos in the BD
geometry. Conclusions are drawn in Sect. 4.

\section{Static Spherically Symmetric Field in BD Theory}
\setcounter{equation}{0}

Variation of the action (\ref{1.1}) with respect to the tensor
metric $g_{\mu\nu}$ and the scalar field $\phi$ yields to the
field equations \cite{brans}
\begin{equation}\label{2.1}
  R_{\mu\nu}-\frac{1}{2}\, R=\frac{8\pi }{c^4\phi}\, T_{\mu\nu}+
  \frac{\omega}{\phi^2}\left(\phi_{,\mu}\phi_{,\nu}-\frac{1}{2} g_{\mu\nu}
  \phi_{,\alpha}\phi^{,\alpha}\right)+\frac{1}{\phi}\,(\phi_{,\mu ;\nu}-g_{\mu\nu}\Box \phi)
\end{equation}
 for the geometric part, and
\begin{equation}\label{2.2}
  \frac{2\omega}{\phi}\,\Box \phi-\frac{\omega}{\phi^2} \phi_{,\mu}\phi^{,\mu}
  +R=0
\end{equation}
 for the scalar field. $\Box$ is the usual d'Alembert operator in curved space--time and
$T_{\mu\nu}$ is the momentum--energy tensor of matter. The line
element describing a static and isotropic geometry is expressed as
\begin{equation}\label{2.3}
  ds^2=-e^{2\alpha}dt^2+e^{2\beta}[dr^2+r^2(d\theta^2+\sin^2\theta d\varphi^2)]\,,
\end{equation}
 where the functions $\alpha$ and $\beta$ depend on the radial coordinate $r$. The
general solution in the vacuum is given by
\begin{eqnarray}
 e^{2\alpha} & = & e^{2\alpha_0}\left[\frac{1-B/r}{1+B/r}\right]^{2/\lambda}\,,\label{2.4} \\
 e^{2\beta} & = & e^{2\beta_0}\left(1+\frac{B}{r}\right)^4
 \left[\frac{1-B/r}{1+B/r}\right]^{2(\lambda -C-1)/\lambda}\,, \label{2.5} \\
 \phi & = & \phi_0\left[\frac{1-B/r}{1+B/r}\right]^{-C/\lambda} \,, \label{2.6}
\end{eqnarray}
 where the constants, appropriately chosen, are given by
\begin{equation}\label{2.7}
  \lambda =\sqrt{\frac{2\omega +3}{2(\omega +2)}}\,, \quad C\cong -\frac{1}{2+\omega}\,,
  \quad \alpha_0=0=\beta_0\,,
\end{equation}
 $$
 \phi_0=\frac{4+2\omega}{G_N(3+2\omega)}\,, \quad B=\frac{M}{2c^2\phi_0}\,
 \sqrt{\frac{2\omega+4}{2\omega+3}}\,.
 $$
 In the weak field approximation, the components of the tensor metric, $g_{\mu\nu}
\simeq \eta_{\mu\nu}+h_{\mu\nu}$, reduces to the form \cite{brans}
\begin{eqnarray}
  g_{00} & \simeq & -1+\frac{2M\phi_0^{-1}}{c^2 r}\frac{4+2\omega}{3+2\omega}\,,
  \label{2.8}\\
  g_{ii} & \sim &  1+\frac{2M\phi_0^{-1}}{c^2 r}\frac{2+2\omega}{3+2\omega}\,,
  \quad i=1,2,3, \label{2.9} \\
  g_{0i} & = & 0\,, \qquad g_{ij}=0\,,\quad i\neq j\,, \label{2.10}\\
  \phi & = & \phi_0 +\frac{2M}{c^2 r}\frac{1}{3+2\omega} \label{2.11} \,.
\end{eqnarray}
As discussed in Introduction, the weak--field solutions
(\ref{2.8})--(\ref{2.11}) have been analyzed for gravitational
red--shift, the deflection of light and perihelion of Mercury (to
be more precise, the last one requires an approximation up to the
second order in $M/r$). In the next Section we will investigate
the phenomenological consequences of BD solutions
(\ref{2.8})--(\ref{2.10}) on neutrinos propagating in such a
geometry.

\section{Neutrino Oscillations in BD Geometry}
\setcounter{equation}{0}

The effects of gravitational fields on the quantum mechanical
neutrino oscillation phases have been analyzed in the
semi--classical approximation, in which the action of a particle
is considered as a quantum phase \cite{stodolsky}. In calculating
such effects induced by BD geometry, we will use the same
approximation.

A particle propagating in a gravitational field from a point A to
a point B, changes its quantum mechanical phase according to the
relation \cite{stodolsky}
\begin{equation}\label{3.1}
\Phi =\frac{1}{\hbar}\int_A^B m ds=\frac{1}{\hbar}\int_A^B
p_{\mu}dx^{\mu}\,,
\end{equation}
where $p_{\mu}=mg_{\mu\nu}(dx^{\nu}/ds)$ is the four--momentum of
the particle and $ds^2=g_{\mu\nu}dx^{\mu}dx^{\nu}$. Following Ref.
\cite{giunti}, the quantum mechanical phase becomes
\begin{equation}\label{3.2}
\Phi =\frac{1}{\hbar}\int_{r_A}^{r_B}\left[E\frac{dt}{dr}-
 p_{r}\right]dr\,.
\end{equation}
Inserting the momentum of the particle, coming from the
shell--condition, $g_{\mu\nu}p^{\mu}p^{\nu}=m^2$,
\begin{equation}\label{3.3}
 p_{r}=e^{\beta-\alpha}\,\sqrt{E^2-m^2e^{2\alpha}}
\end{equation}
into Eq. (\ref{3.2}), and using the fact that
$dt/dr=e^{\beta-\alpha}$, one gets a difference phase $\Delta\Phi$
given by
 \begin{equation}\label{3.4}
 \Delta\Phi =\frac{\Delta}{2\hbar E}\, \int_{r_A}^{r_B}
 \left(1+\frac{B}{r}\right)^2\left(\frac{1-B/r}{1+B/r}\right)^{(\lambda -C)/\lambda}
 \,\left[E-\sqrt{E^2-m^2\left(\frac{1-B/r}{1+B/r}\right)^{2/\lambda}}\right]\,dr\,.
 \end{equation}
By using the weak field approximation, Eqs.
(\ref{2.8})--(\ref{2.10}), one can separate out the {\it
gravitational} contribution to the neutrino oscillation phase, so
that Eq. (\ref{3.4}) can be cast in the form
\begin{equation}\label{3.5}
 \Delta\Phi=\Delta\Phi_{0}+\Delta\Phi_{\omega}\,,
\end{equation}
where (restoring the constants $c$ and $\hbar$)
\begin{equation}\label{3.6}
 \Delta\Phi_{0}=\frac{\Delta m^2 c^3}{2E\hbar}(r_B-r_A)\,,
\end{equation}
which represents the standard phase of neutrino oscillations, and
\begin{equation}\label{3.7}
 \Delta\Phi_{\omega}=\frac{\Delta m^2 c}{2\hbar E}\frac{G_NM}{2+\omega}\log\frac{r_B}{r_A}\,{.}
\end{equation}
In deriving Eqs. (\ref{3.6}) and (\ref{3.7}), we have considered
ultra--relativistic neutrinos, $E>>m$, where $E$ is interpreted as
the energy at the infinite (see \cite{giunti} for details). The
integration has been performed along the light--ray trajectory
where $E$ is constant.

It is convenient to rewrite the phases (\ref{3.6}) and (\ref{3.7})
in the following way
\begin{equation}\label{3.8}
\Delta\Phi_{0}\approx 2.5\cdot 10^3 \frac{\Delta
m^2}{\mbox{eV}^2/c^4}\,\frac{\mbox{MeV}}{E}\,\frac{r_B-r_A}{\mbox{Km}}\,{,}
\end{equation}
and
\begin{equation}\label{3.9}
 \Delta\Phi_{\omega}\approx 3.5\cdot 10^3\, \frac{1}{2+\omega}\,
 \frac{\Delta m^2}{\mbox{eV}^2/c^4}\,
 \frac{\mbox{MeV}}{E}\,\frac{M}{M_{\odot}}\, \log\frac{r_B}{r_A}\,{,}
\end{equation}
where $M_{\odot}$ is the solar mass. Estimations of the difference
phases (\ref{3.8}) and (\ref{3.9}) are carried out for solar,
atmospheric and astrophysical neutrinos. To this end, we will
introduce the ratio $q$ defined as
\begin{equation}\label{3.10}
  q=\frac{\Delta\Phi_{\omega}}{\Delta\Phi_0}\approx 1.5 \, \frac{1}{2+\omega}\,
  \frac{M}{M_{\odot}}\, \frac{\log (r_B/r_A)}{(r_B-r_A)/\mbox{Km}}\,.
\end{equation}
 $q$ does not depend on the squared--mass difference $\Delta m^2$ and on
the neutrino energy $E$. For solar neutrinos, we use the following
values: $M\sim M_{\odot}$, $r_A\sim r_{Earth}\sim 6.3\cdot
10^3$Km, and $r_B\sim r_A+D$, where $D\sim 1.5\cdot 10^8$Km is the
Sun--Earth distance. Eq. (\ref{3.10}) gives the result
\begin{equation}\label{3.11}
  q\sim 10^{-8}\, \frac{1}{2+\omega}\,,
\end{equation}
 which is an irrelevant correction to the difference phase (\ref{3.8}). Analogous
conclusion holds for atmospheric neutrinos.

Concerning the astrophysical neutrinos, the effect could be more
relevant and could be measured by terrestrial experiments. In
fact, setting $r_B=\alpha r_A$, $1< \alpha \leq \infty$ and using
the typical values of neutron stars, $M\sim 1.4 M_{\odot}$ and
radius $r_A\sim 10$Km as in Ref. \cite{burgard}, we get
\begin{equation}\label{3.12}
  q\sim \frac{0.2}{2+\omega}\, \frac{\log \alpha}{\alpha -1}\,.
\end{equation}
 Till now, our analysis has been done for radially propagating neutrinos.
In the case of motion transverse to the radial propagation and
near to the detection point $r_A$ we have, following
\cite{burgard},
\begin{equation}\label{3.13}
\Delta\Phi_{\omega}^{\bot}=\frac{\Delta m^2 c}{2\hbar
E}\frac{G_NM}{2+\omega}\frac{r_B-r_A}{r_A}\approx
 3.5\cdot 10^3\,\frac{1}{2+\omega}\, \frac{\Delta m^2}{\mbox{eV}^2}\,
 \frac{\mbox{MeV}}{E}\,\frac{M}{M_{\odot}}\, \frac{r_B-r_A}{r_A}\,{.}
\end{equation}
Then, the ratio between the difference phases (\ref{3.13}) and
(\ref{3.8}) is
\begin{equation}\label{3.14}
  q^{\bot}=\frac{\Delta\Phi_{\omega}^{\bot}}{\Delta\Phi_0}\approx 1.5\, \frac{1}{2+\omega}\,
  \frac{M}{M_{\odot}}\, \frac{\mbox{Km}}{r_A}\,.
\end{equation}
For the numerical constants corresponding to Sun and Earth, we
have
\begin{equation}\label{3.15}
  q^{\bot}_{Sun}\sim \frac{1.5\cdot
  10^{-5}}{2+\omega}\,,\qquad
   q^{\bot}_{Earth}\sim \frac{5\cdot
  10^{-10}}{2+\omega}\,.
\end{equation}
Using the above values for a neutron star, Eq. (\ref{3.14}) gives
the result
\begin{equation}\label{3.16}
 q^{\bot}\sim \frac{0.2}{2+\omega}\,.
\end{equation}
As discussed in Introduction, experimental data imply that the
parameter $\omega$ can assume the value $\omega\geq 500$. For the
lower limit, one gets from Eqs. (\ref{3.12}) and (\ref{3.16}),
\begin{equation}\label{3.17}
  q\sim 4\cdot 10^{-4}\, \frac{\log \alpha}{\alpha -1}\,, \qquad
  q^{\bot}\sim 4\cdot 10^{-4}\,,
\end{equation}
giving a correction of the $0.01$ percent.
 Values of $\omega\leq 30$, coming from the anisotropy of microwave background
radiation, allow to get corrections of few percents, as one can
immediately derive from Eqs. (\ref{3.12}) and (\ref{3.16}). Such
contributions to the quantum mechanical phase of neutrinos, are
very significant and could be considered as a test for
establishing the validity of BD theory.

\section{Conclusions}

In this paper, we have analyzed neutrino oscillation phenomena in
the framework of BD theory. We have derived a correction to the
standard difference phase of the order $G_N\Delta m^2$, which
vanishes in the limit $\omega\to \infty$, when the BD theory
reduces to General Relativity.

Estimation of such a correction has been carried out assuming for
the parameter $\omega$ the values $\omega\sim 500$ and $\omega\leq
30$. Such values may be relaxed considerably with the advances in
technology associated with astronomical observations and
astrophysical experiments, making our corrections as a mean to
discern between BD theory and Einstein's theory, in addition to
the ones discussed in the Introduction.

Nevertheless, BD is a particular case of scalar tensor--theories
where one assumes that matter acts as source of scalar field
$\phi$, which generates the curvature of space--time associated to
the metric. The strength of the coupling between the scalar field
and gravity is determined, in these theories, by the function
$\omega (\phi )$, which is constant in the BD theory. Besides, a
self--interaction potential $V(\phi)$ can be also introduced,
generalizing in such a way dynamics of the field.

The dependence of the parameter $\omega$ on $\phi$ could have the
property that, at the present epoch, and in weak field situations,
the value of the scalar field $\phi_0$ is such that $\omega$ is
very large, leading to theories almost identical to General
Relativity today, but for past or future values of $\phi$, as in
strong field regimes as for neutron stars, $\omega$ could take
values that would lead to significant differences from General
Relativity. In this sense, scalar--tensor theories are richer than
BD theory and could play a relevant role in the neutrino
oscillation physics (and in Pound--Rebka or COW experiments, as
well as in atomic systems in linear superposition of different
energy eigenstates). This because the variability of the parameter
$\omega$ implies that, in some epoch, its value could be very
small, and, in such a way, a correction to the quantum mechanical
phase of $10\%$ can be obtained (in this particular case, the
factor (\ref{1.4}) reduces the Ahluwalia--Burgard result to
$15\%$, instead of $20\%$ as derived in Ref. \cite{burgard}). In a
forthcoming paper we will face these issues.

\vspace{0.3cm}

The authors would like to thank the referees, and in particular D.V Ahluwalia, for the
useful comments on the subject treated in this paper.

\end{document}